**Automated cantilever exchange and optical alignment for High-throughput, parallel atomic force microscopy**

Tom Bijnagte, Geerten Kramer, Lukas Kramer, Bert Dekker, Rodolf Herfst, Hamed Sadeghian[1]

Department of Optomechatronics, Netherlands Organization for Applied Scientific Research, TNO, Delft, The Netherlands

[1] Hamed.sadeghianmarnani@tno.nl

**Abstract**

In atomic force microscopy (AFM), the exchange and alignment of the AFM cantilever with respect to the optical beam and position-sensitive detector (PSD) are often performed manually. This process is tedious and time-consuming and sometimes damages the cantilever or tip. To increase the throughput of AFM in industrial applications, the ability to automatically exchange and align the cantilever in a very short time with sufficient accuracy is required. In this paper, we present the development of an automated cantilever exchange and optical alignment instrument. We present an experimental proof of principle by exchanging various types of AFM cantilevers in 6 seconds with an accuracy better than 2 µm. The exchange and alignment unit is miniaturized to allow for integration in a parallel AFM. The reliability of the demonstrator has also been evaluated. Ten thousand continuous exchange and alignment cycles were performed without failure. The automated exchange and alignment of the AFM cantilever overcome a large hurdle toward bringing AFM into high-volume manufacturing and industrial applications.

**I. Introduction**

In atomic force microscopy (AFM), a micro cantilever with a sharp tip is scanned over the surface of a sample. The motion of the cantilever is measured using optical beam deflection (OBD),[1] which is converted to a topography measurement of the surface.[2,3] Due to its atomic resolution and the ability to measure mechanical,[4] physical[5] and chemical[6] parameters, AFM has attracted great interest in several industrial applications, such as semiconductor metrology,[7,8] biological and medical applications,[5] material science[9] and data storage.[10] Industrial applications with high-volume manufacturing characteristics require high-throughput, fully automated instruments. Automation is required to minimize user input and measurement time. Substantial efforts have been made toward automated positioning and alignment of the sample toward the AFM tip and image acquisition.[11,12] The AFM tip must be replaced when the tip becomes blunt. Moreover, after the cantilever is exchanged, the OBD must be aligned with respect to the cantilever. This process is typically performed in two steps: 1) the laser spot is located at the end of the cantilever to maximize the signal-to-noise ratio and 2) the spot is aligned in the center of the position-sensitive detector (PSD). The manual exchange of an AFM tip and OBD alignment are tedious, time-consuming procedures. For industrial applications, to keep the throughput of the measurement high, it is very important that the cantilever exchange and OBD alignment be performed automatically. Even for research applications, scientists working with AFM would appreciate an automated probe exchange and alignment, which would save a substantial amount of time.

The state-of-the-art automatic cantilever exchange is performed using either magnetic actuation[13] or a vacuum[14] with an intermediate holder. For the magnetic exchange mechanism, a magnetic layer with alignment grooves must be added to the AFM cantilever's chip. The use of an intermediate holder results in an additional mass on the z-scanner, which reduces the resonance frequency of the z-stage and thus the bandwidth of the measurement.[15]

Furthermore, the state-of-the-art exchange and alignment procedure takes from 1 minute up to 3 minutes.

In this paper, we present the development of a rapid, accurate and fully automated cantilever exchange followed by an automatic optical alignment. The instrument for this procedure is specifically designed for use in the recently developed parallel AFM.[2,7,16] The automatic cantilever exchange and alignment instrument is capable of exchanging 22 cantilevers in parallel in only 6 seconds and includes OBD alignment for a high signal-to-noise ratio (SNR). No intermediate holder is used, maximizing the performance of the z-scanner and rendering the method compatible with any type of AFM cantilever.

This paper is organized as follows. In the following section, the functions and requirements are defined, and the conceptual architecture of the automated cantilever exchange and alignment instrument is described. The third section contains the detailed design and realization of the instrument. The experimental results of the exchange and alignment, their reproducibility and, finally, the AFM imaging performance are presented in section 4. Conclusions are provided in Section 5.

## II. System requirements and conceptual architecture

The automated cantilever exchange system was developed to meet four primary system requirements: short exchange time, alignment accuracy, reliability, and cantilever chip clamping. The required exchange time is derived from the required overall throughput in the parallel AFM system,[17,18] aiming at a throughput of 20 wafers per hour, or a 180-s cycle time. The cycle includes loading and unloading (6 seconds), scanning (160 seconds for a field of 10 µm×10 µm), pickup and positioning of the miniaturized AFM (MAFM) (12 seconds), and wafer exchange and alignment (6 seconds and 2 seconds, respectively). To further enhance the throughput of the parallel AFM, the cantilever exchange and alignment are envisioned to be performed in parallel with the wafer exchange, within 6 seconds.

The required alignment accuracy is derived from the required SNR of the OBD and the PSD alignment. The MAFM[2,19] uses an OBD containing a laser beam reflecting the cantilever to a position-sensitive detector (PSD).[1] After a new tip is exchanged, the optical beam to the PSD must meet the following requirements. To attain a sufficiently high SNR, the minimal required output sum signal from the PSD is 0.7 V. For sufficient sensitivity, the top-bottom signal must lie between -1.5 V and +1.5 V.

The reliability, repeatability and reproducibility of the entire cantilever exchange and alignment process are very important for industrial application. A typical acceptable reliability is 1 operator assistance per week, based on 24/7 service. For 44 MAFMs in the parallel AFM,[2] at 20 wafers per hour and a tip life-time of 1 wafer, the mean time between operator assistance is 150,000 hours. Therefore, the reliability target for the demonstrator is set to 10,000 cycles without operator interruption. Another important function of the cantilever exchange is to automatically fix and release a cantilever on and from the MAFM. During the scanning function, the cantilever chip is imposed to dither accelerations and vertical scan accelerations. Clamping errors directly translate into visible image distortions, such as drift in the horizontal direction and noise in the vertical direction. For a scan size of 10×10 µm$^2$, we set 1% or 40 nm as an acceptable drift requirement in the horizontal direction and 0.1 nm for vertical clamp induction. Sufficient force should be generated to ensure that the cantilever stays in place under the scanning accelerations.

To meet the aforementioned requirements on the cantilever exchange, OBD alignment and cantilever clamp on the MAFM, several functions are considered, as follows:

1. Storage of new cantilevers and used cantilevers in a stock

2. Pick-up of the cantilevers from storage

3. Transfer of the cantilever from storage to the MAFM

4. Take-over of the cantilever from the exchange unit to the MAFM

5. Positioning of the cantilever with respect to the OBD

6. Clamping of the aligned cantilever to the MAFM

The concept architecture will be discussed in the following sub-section based on the above system functions.

**A. Concept architecture**

An important criterion for the concept architecture is the compatibility of the exchange system with any cantilever type without modification of its chip. Any modification to the chip requires an additional process step in the AFM chip manufacturing, resulting in a higher cost of consumables. Moreover, miniaturizing the cantilever exchange is also very important to parallelize the cantilever exchange process for high throughput in a parallel AFM.[2,18]

The cantilever exchange was implemented in the form of a long stroke conveyor robot (LoS-Y), moving over the 22 MAFMs, provided with 22 short stroke xyz-stages with a pitch of 20 mm, equal to the pitch of the MAFMs,[2,17] referred to as the Chip Manipulator(CM), see FIG 1. The conveyor robot enables the transfer of each cantilever among stock 1, the MAFM and stock 2. Approaching from above, the CM picks up the chip from the stock. The stock capacity is 5 probes per MAFM or 220 probes for the entire demonstrator. With all of the CMs in the upper position, the LoS-Y moves the probes to a position above their corresponding MAFMs.

Each CM module is equipped with two vacuum pads for holding the two cantilever's chips. This step enables new chip preparation at one CM slot parallel to the MAFM positioning at the exchange position. As soon as the MAFM is ready for exchange, the old chip is removed and placed at the remaining free slot at the CM. The conveyor robot moves further, and the new chip is aligned and placed at the MAFM. Next, the old chip is moved to the output stock parallel to the MAFM positioning before scanning.

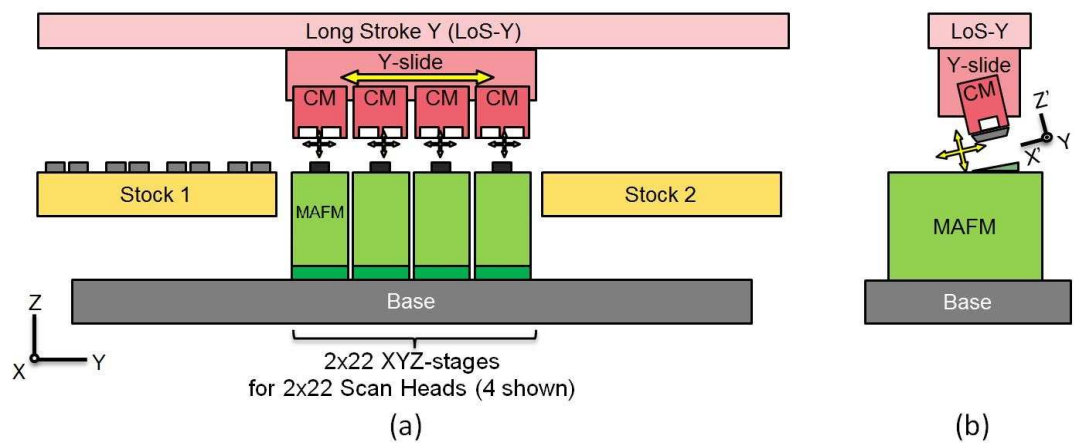

FIG 1. Schematic illustration of the cantilever exchange demonstrator, showing the (a) front view and (b) side view. All of the tips can be exchanged in parallel within 6 s. Equipping the exchange robot with 1, 2, 4, or 11 CM modules will reduce the tool's cost against an increased exchange time.

The cantilever's alignment function was implemented using a miniaturized xy-stage, scanning the cantilever across the OBD (fast in the lateral direction and slow in the longitudinal direction), as depicted in FIG. **2**. Therefore, each CM is equipped with a fast xy-stage with closed-loop position feedback. The position of the cantilever with respect to the OBD is measured by the PSD's output. When the cantilever is outside the OBD, no light is reflected to the PSD. When the cantilever enters the beam (as shown in Figure 2b and 2c), a portion of the light is reflected to the PSD, and the first edge is detected. Moving on, the full beam is reflected up to the second cantilever edge. A decreased intensity on the PSD indicates the position of the second edge. The final aligned position in the lateral direction is the center between the two stage positions at which the two edges were detected. The longitudinal position is found by moving the cantilever along its center line over the OBD. A rapidly decreasing light intensity at the PSD defines the aligned position at the end of the cantilever.

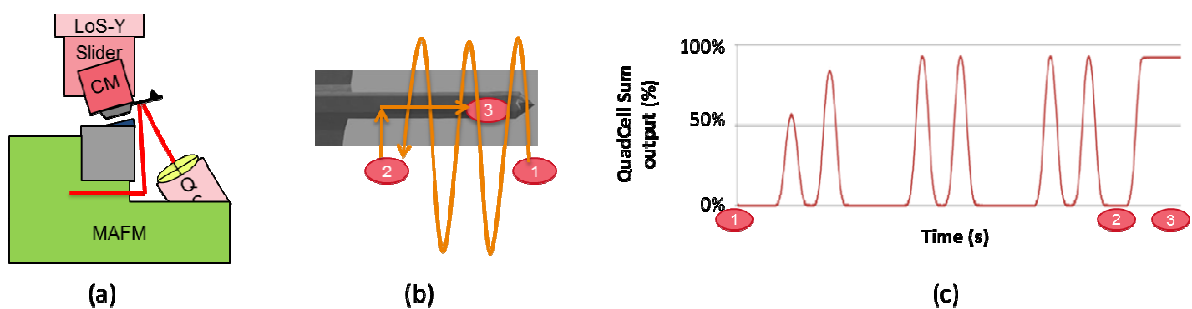

FIG. 2. Schematic illustration of the cantilever alignment, showing (a) the probe chip at the CM in the aligned position, (b) a typical alignment motion, first scanning to find the first edge of the cantilever (schematically shown as sinusoidal motion) and then moving along the cantilever center to the tip, and (c) the PSD output time-log corresponding to the typical alignment motion.

After the cantilever is aligned with respect to the OBD, the cantilever should be taken over by the MAFM. Any position deviations during this take-over will disturb the OBD position at the end of the cantilever.

The gap at which the take-over occurs determines the stability of the alignment during the take-over. A measurement of the gap during the take-over is provided from the varying air flow and pressure during the approach, as illustrated in FIG. **3**. The MAFM starts in its lowest position, while the CM starts at the pre-calibrated take-over position. First, the MAFM vacuum clamp is switched on, and the ambient pressure $p_{amb}$ is read from a pressure sensor. Next, the MAFM moves up, causing the gap between the probe and MAFM clamp to decrease, which induces a higher pressure drop over the gap. Once the pressure sensor reads the pre-calibrated take-over pressure level $P_{TO}$, the intended gap height is reached, and the MAFM stops.

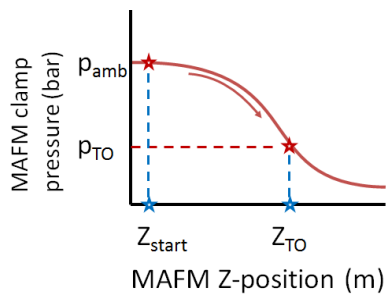

FIG. 3. Illustration of the MAFM clamp pressure as a function of the MAFM z-position.

The take-over occurs without any force required between the cantilever's chip and the MAFM. With the beam aligned at the cantilever's center, the vacuum clamp at the CM is vented; simultaneously, the vacuum at the MAFM clamp is evacuated, causing the chip to fly over from the CM clamp to the MAFM clamp.

Unlike the state-of-the-art approach used in clamping the cantilever to the MAFM, which requires an intermediate holder, the clamping at the MAFM is achieved with a vacuum clamp on top of the dither piezo.[2] The advantage of this technique over other clamp methods[13] is that the vacuum clamp does not add mass to the moving parts of the vertical (z) scanning stage; therefore, the clamp will not limit the vertical (z) scanning stage control bandwidth. Furthermore, the vacuum principle allows for a compact design, which is more difficult to achieve with mechanical clamp solutions. The vacuum clamp area is approximately 1.9 mm$^2$. Applying a normal industrial vacuum level of -0.7 bar with reference to the ambient pressure leads to a preload force of 0.13 N. Given the 3.2-mg mass of a chip, vertical accelerations of up to 42 km/s$^2$ are feasible, divided into 8 km/s$^2$ for vertical (z) scanning up to an amplitude of 0.5 µm at 20 kHz and 34 km/s$^2$ for the cantilever's excitation accelerations up to an amplitude of 0.3 nm at 1.7 MHz.

## III. Detailed design and realization

In this section, the detailed design and realization of the demonstrator, including the CM design, motion control architecture, and software architecture, are presented.

The CM module picks up the new cantilever from stock 1, removes the old cantilever from the MAFM, and positions the new cantilever in the x-, y-, and z-direction with respect to the OBD. The required position accuracy is 1 µm (3σ) in the horizontal direction and 2 µm (3σ) for the vertical direction. The required time to the target, including the settling time, is 0.3 s or less.

The core of the CM module is a horizontal 2D stage. As depicted in FIG. **4**(a), the actuation and guiding in the x-direction are provided by a set of two parallel bender piezo-electric elements. The deflection of these benders is measured using two strain gauges. The first set of benders is connected to an identical second set, which is mounted perpendicularly to the first set, thus providing the actuation and guiding in the y-direction. Two vacuum pads are integrated in the xy-stage and are supplied via two flexible hoses.

The entire xy-stage is suspended using flexures, which guide the stage module in the vertical (z) direction. A piezo stepper motor actuates the stage in this direction, while the actual position is measured with a linear encoder. Additionally, the piezo stepper motor is suspended in elastic guides, which are preloaded for a stiff connection between the motor and the CM base. In the case of an accidental collision with the stock or MAFM, the motor will disconnect at a safe force level, thus preventing the probe chip from breaking. FIG. **4**(b) depicts a computer-aided design (CAD) of the complete CM module.

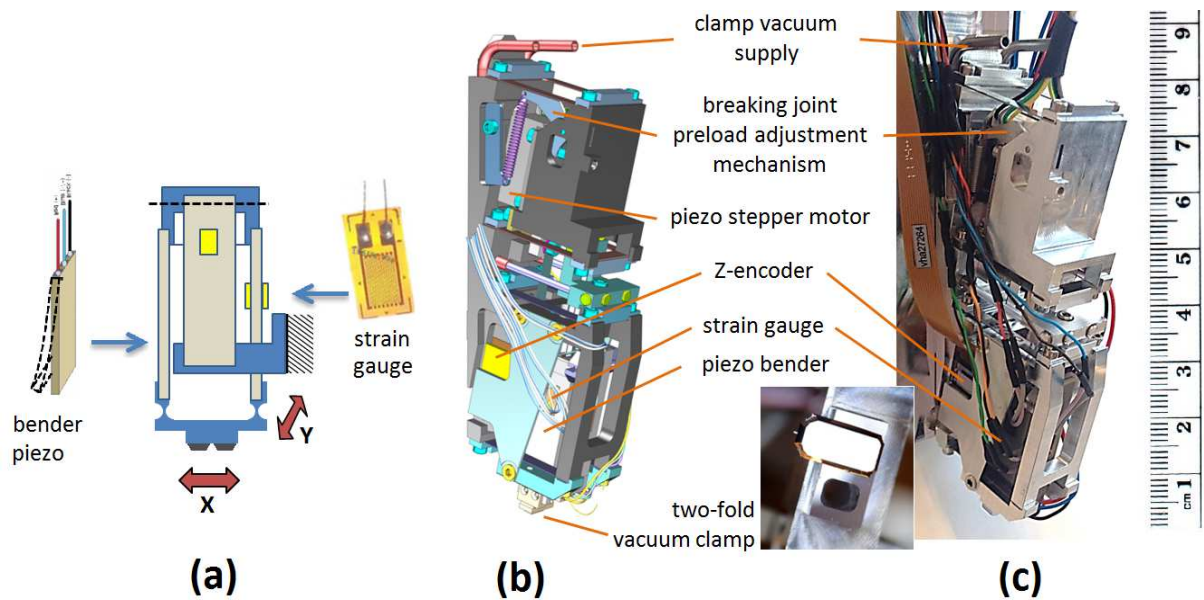

FIG. 4. Detailed CM design, showing (a) a schematic representation of the short stroke xy-stage, actuated via bender piezo elements, (b) a CAD image, and (c) a photograph of the realized module and a magnified view of a probe chip at the vacuum clamp.

FIG. **4**(c) shows the realized CM module. A test setup was built to verify the CM performance with respect to the chip's clamping, moving, and positioning functions.

The clamping functionality of the CM was evaluated by a 24-hour endurance test, with continuous scanning at a frequency of 25 Hz over the full stroke. A comparison of close-up pictures before and after the endurance test revealed no visible difference in the cantilever chip displacement.

The accurate movement and positioning of the cantilever by the CM was demonstrated by moving the stage over a pre-defined distance while measuring the stage position error. The results are shown in FIG. **5**. FIG. **5**(a) shows the position error of the y-stage, which is the most critical horizontal motion because it is the fast motion for scanning. The achieved positioning time for a 300-µm step, including settling, is 0.18 s, less than the required 0.3 s. The position noise after settling is 0.02 µm (3σ).

As shown in FIG. **5**(b), the time required for vertical positioning by the CM (over a 1.5-mm step), including settling, is 0.25 seconds, less than the required 0.3 s. The position noise after settling is 0.02 µm (3σ).

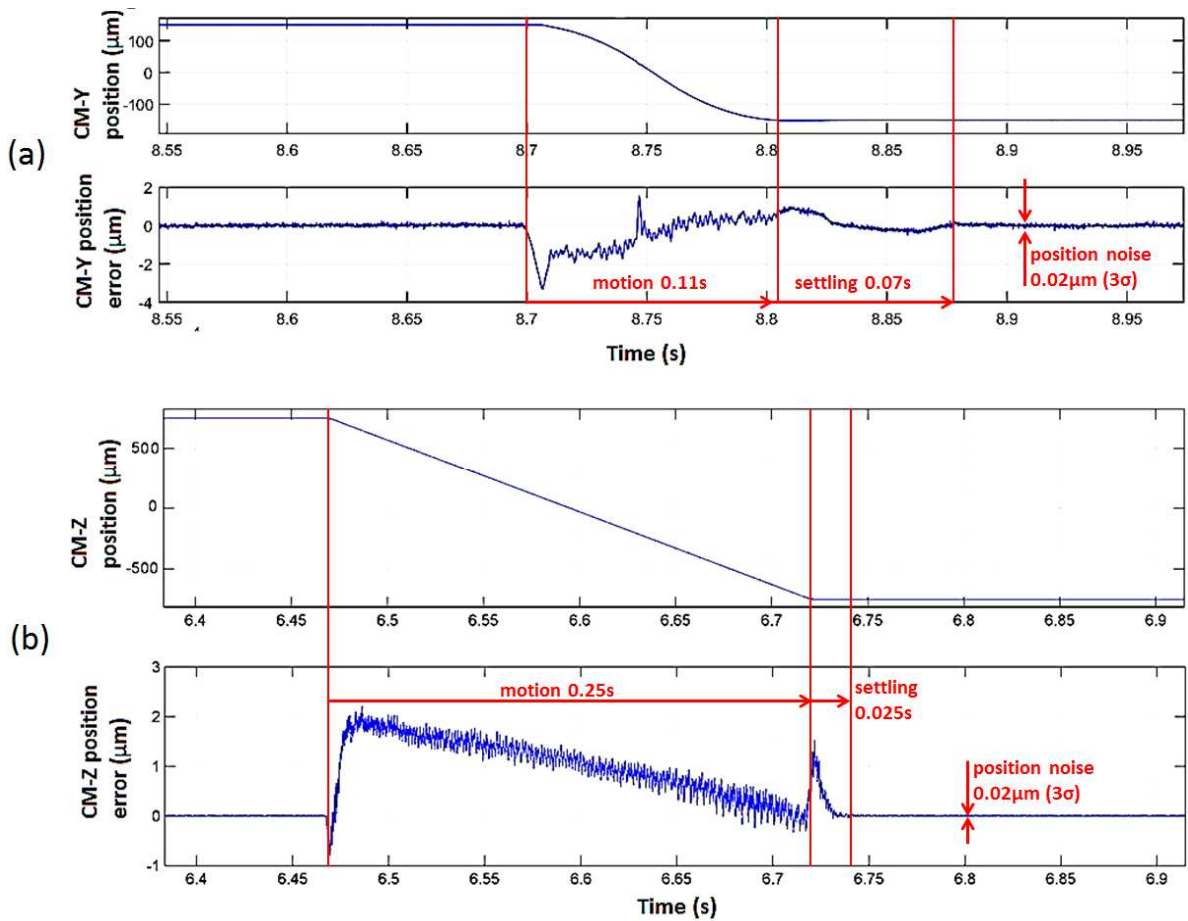

FIG. 5. Illustration of the CM motion, displaying (a) horizontal CM motion, showing the position and position error in the rapid scanning y-direction over a 300-μm horizontal step, and (b) vertical CM motion, showing the position and position error over a 1.5-mm vertical step.

For the system control, a dSpace system was used. This system contains a flexible programmable real-time processor that runs all of the safety and motion control tasks. It is driven and programmed from an operator PC running Windows and MATLAB®. The real-time processor is programmed using MATLAB® Simulink®, and for the logic, MATLAB® Stateflow® is used. The position actuators, position sensors and pressure sensors in the system are directly connected to the dSpace interface boards, as depicted in FIG. **6**(a).

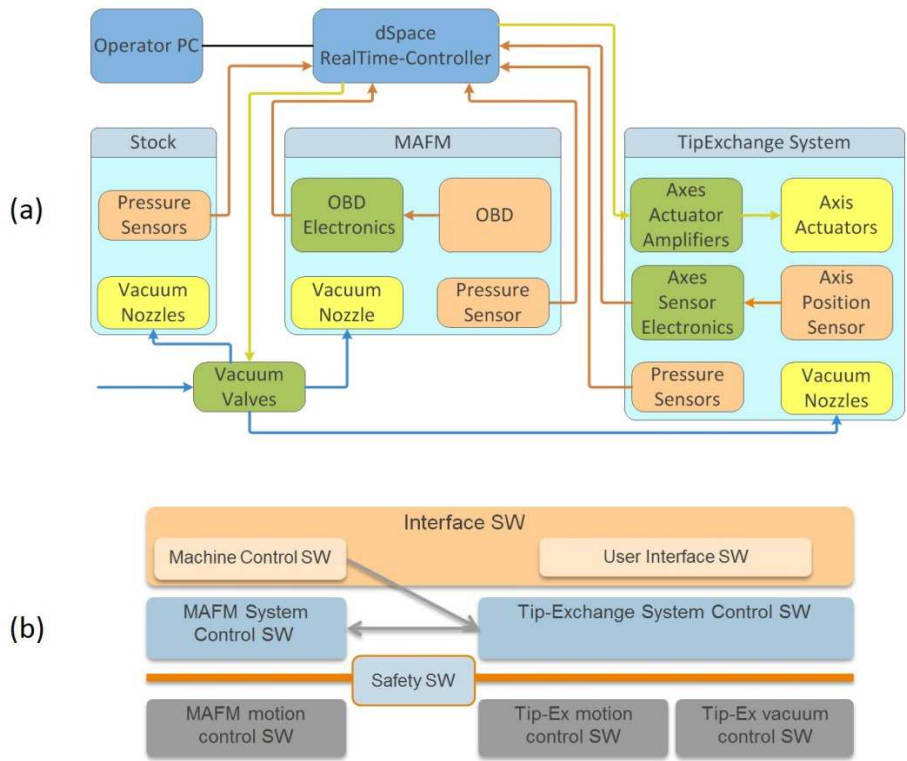

FIG. 6: Illustration of the control system, showing (a) a system connection diagram and (b) a schematic representation of the system control software functions.

 The Los-Y actuator is a linear motor, dimensioned to the force required to accelerate the mass of all 22 parallelized CMs over the required 500-mm travel range, within the allowable time of 0.6 s. A Technotion® actuator, driven by a Trust linear amplifier was chosen to satisfy the requirements. The control design for this axis is a standard PID controller, which is implemented as a discrete state-space design. The achieved control bandwidth is 235 Hz.

 A piezo walker from PiezoMotor® (Piezo LEGS® Linear Twin 20N) was chosen for the CM z-actuation. This actuator is sufficiently small, can deliver an adequate amount of force (15 N), and can also reach the required speed (9 mm/s). Because this actuator is a piezo-walker, its reproducibility is not sufficient to reach the required absolute positioning accuracy of less than 2 µm (3σ). The required accuracy was achieved using a small chip encoder, read through the analog sin-cos encoder interface of the DS3002 board. The gain fluctuation of this actuator offers some spread in the control bandwidth. The CM z-actuator is used in a closed loop (speed control) and achieves a bandwidth of 23-60 Hz (gain fluctuation of approximately 6 dB).

The core of the software functions is the Tip-Exchange System Control Software, as shown in FIG. **6**(b). Based on inputs from the User Interface Software and the overall Machine Control Software, the Tip-Exchange System Control Software initiates the motion of the axes and vacuum valves and reads the vacuum and position sensors to control the higher-level exchange and alignment procedures.

The MAFM Control Software (SW) is part of the overall control architecture and is separated from the Tip-Exchange System Control Software. To control the MAFM functions, the cantilever Exchange Software interfaces with the MAFM Control SW.

A layer called 'Safety Software' is placed as a barrier between the Tip-Exchange Software and the lower-level Motion Control and Vacuum Control Software. This layer continuously monitors the system state and immediately stops the system from running in situations such as an unexpected loss of vacuum or motion leading to a collision.

## IV. EXPERIMENTAL RESULTS

A PoP setup is built to demonstrate the system requirements: short exchange time, alignment accuracy, reliability, and probe chip clamping. The demonstrator includes one CM, one MAFM and a cantilever stock with five chip pockets to hold different cantilever types. Three different types of commercially available cantilevers are tested to demonstrate the tip-exchange functionality and performance, namely, the Arrow UHF Triangular, Olympus BL-AC40TS Rectangular and Bruker TESPA-V2. FIG. **7** shows the realized demonstrator for cantilever exchange and alignment.

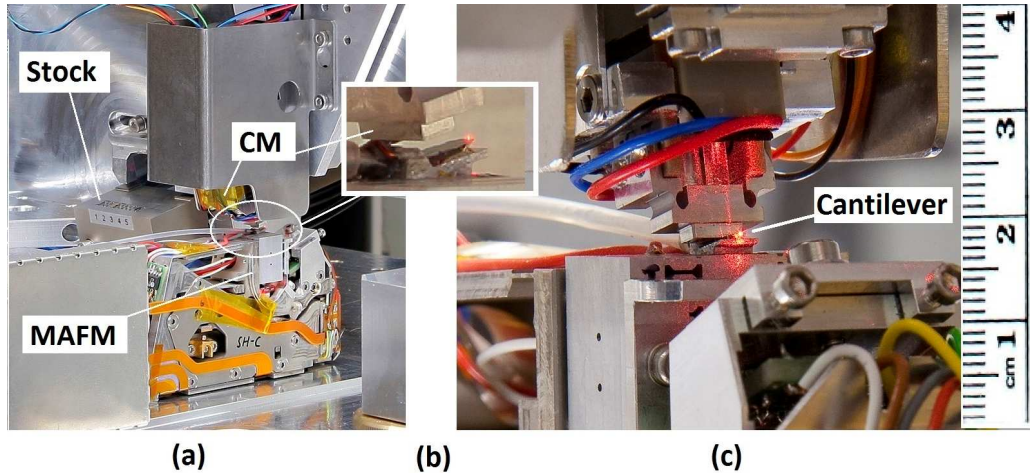

FIG. 7. Photographs of the realized cantilever exchange demonstrator setup, showing (a) the MAFM, CM and stock of cantilevers and (b, c) close ups of an aligned cantilever automatically placed at the MAFM.

For the performance tests, the OBD system in the MAFM is manually calibrated one time for the Olympus BL-AC40TS cantilever, such that the PSD top/bottom output signal is approximately 0 V. The Arrow UHF and Bruker TESPA probes were placed without OBD realignment.

The achievable exchange time was tested. The measured exchange time required to pick up the old cantilever and place a new cantilever on the MAFM is 6.0 ± 0.8 s (3σ), measured during a run of 100 cycles for each cantilever type. These measurement results are shown in FIG 8. This exchange time includes downward movement of the CM, approach of the MAFM, pick up of the old cantilever from the MAFM, alignment of the new cantilever, placement of the new cantilever on the MAFM and movement of the CM to the initial position.

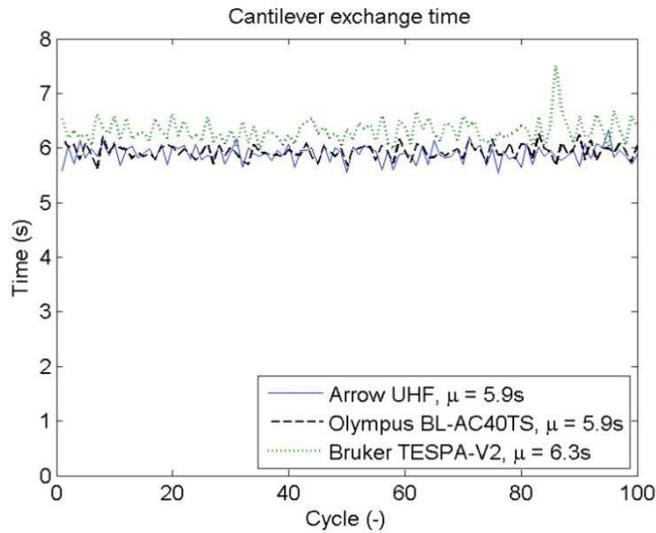

FIG 8. Measured time to pick up an old cantilever and place a new cantilever on the MAFM, for 3 types of cantilevers.

The second test concerns the alignment accuracy. The three cantilevers were successfully placed automatically at the MAFM clamp over 100 cycles for each cantilever type, as shown in FIG. **9**. The maximum reflectance of the OBD beam on the PSD results in a maximum PSD voltage of 1.34 V, 1.15 V and 1.75 V for the Arrow UHF, Olympus BL-AC40TS and Bruker TESPA-V2 cantilevers, respectively. As shown in FIG. **9**, all 300 cantilever exchanges were successfully above the required PSD voltage of 0.7 V.

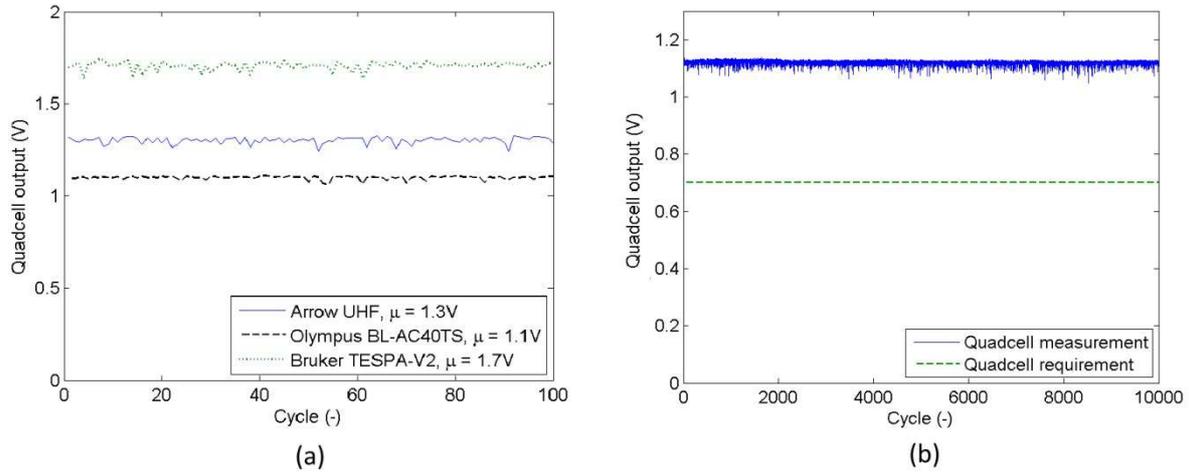

FIG. 9. Measurement results of OBD beam positioning at the PSD after automated cantilever exchange, showing the PSD sum signal over (a) 100 exchange cycles for 3 different cantilever types and (b) 10,000 exchange cycles for two Olympus BL-AC40TS cantilevers. For a sufficient signal-to-noise ratio, the output voltage should be at least 0.7 V.

The third validation test shows the system reliability. The measured PSD sensor read out after cantilever placement over 10,000 cycles is shown in FIG. **9**(b) and indicates that the exchange of the cantilever is accurate and reliable; the output voltage is also above the PSD placement requirement of 0.7 V.

The impact of the new vacuum clamp on the imaging performance is also tested. A high-frequency cantilever (resonance frequency above 1 MHz)—i.e., Arrow UHF, was loaded by the cantilever exchange demonstrator to the MAFM, and a reference image was acquired. For this purpose, a small xy-stage was placed above the MAFM, holding a 5×5-mm sample with a grating pattern. The cantilever was unloaded from the MAFM and placed back in the probe stock, followed by pick up from the stock, alignment on the MAFM, and placement at the MAFM vacuum chuck. This cycle was repeated 150 times, and a final image was taken. FIG. **10** shows the reference and final images.

From FIG. **10**, it can be concluded that, after 150 exchanges, the vacuum clamp still keeps the probe sufficiently stable for AFM scanning.

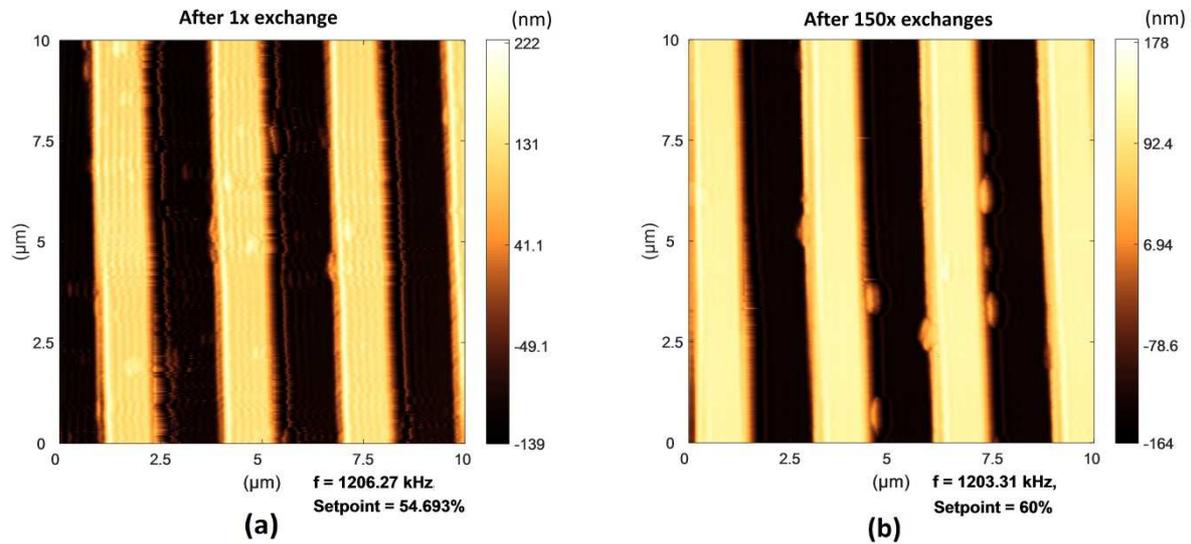

FIG. 10. AFM imaging results (a) after one automatic probe exchange and (b) after 150 automatic exchanges. Both images were acquired from the same TGZ-03 3-micron pitch line sample with a step height of 500 nm. The vertical scale factor was not calibrated.

## V. CONCLUSIONS

In this paper, a proof of principle of automated cantilever exchange, alignment and clamping for atomic force microscopy was demonstrated.

The cantilever exchange functionality, performance and reliability were successfully demonstrated in a single-MAFM demonstrator setup. This automated, miniaturized cantilever exchange and alignment can be integrated in a parallel AFM.[18] The cantilever exchange time is 6 seconds, and the alignment accuracy is better than 2 μm. The accuracy and reliability of the cantilever placement were demonstrated with two successful runs of 5,000 cycles without human interruption.

The imaging performance was verified after 1 and 150 automatic cantilever exchanges. This demonstrator will be further implemented in a parallel AFM demonstrator as a platform for a high-throughput, automated AFM measurement system.

**ACKNOWLEDGMENTS**

This research was supported by the Netherlands Organization for Applied Scientific Research, TNO, Early Research Program 3D Nanomanufacturing Instruments. The research described in this paper was performed in the framework of the SeNaTe project. TNO gratefully acknowledges funding from the ECSEL Joint Undertaking (Grant Agreement no. 662338) and the Netherlands Enterprise Agency RVO.

[1]R. W. Herfst, W. A. Klop, M. Eschen, T. C. van den Dool, N. B. Koster, and H. Sadeghian, "Systematic characterization of optical beam deflection measurement system for micro and nanomechanical systems," Measurement **56**, 104–116 (2014).

[2]H. Sadeghian, R. Herfst, J. Winters, W. Crowcombe, G. Kramer, T. van den Dool, and M. H. van Es, "Development of a detachable high speed miniature scanning probe microscope for large area substrates inspection," Rev. Sci. Instrum. **86**, 113706 (2015).

[3]R. Sri Muthu Mrinalini, R. Sriramshankar, and G. R. Jayanth, "Direct measurement of three-dimensional forces in atomic force microscopy," IEEE/ASME Trans. Mechatron. **20**, 2184–2193 (2015).

[4]M. Favre, J. Polesel-Maris, T. Overstolz, P. Niedermann, S. Dasen, G. Gruener, R. Ischer, P. Vettiger, M. Liley, H. Heinzelmann, and A. Meister, "Parallel AFM imaging and force spectroscopy using two-dimensional probe arrays for applications in cell biology," J. Mol. Recognit. **24**, 446–452 (2011).

[5]A. F. Guedes, F. A. Carvalho, I. Malho, N. Lousada, L. Sargento, and N. C. Santos, "Atomic force microscopy as a tool to evaluate the risk of cardiovascular diseases in patients," Nat. Nanotechnol. **11**, 687–692 (2016).

[6]R. A. Oliver, "Advances in AFM for the electrical characterization of semiconductors," Rep. Prog. Phys. **71**(7), 076501 (2008).

[7]H. Sadeghian, B. Dekker, R. Herfst, J. Winters, A. Eigenraam, R. Rijnbeek, and N. Nulkes, "Demonstration of parallel scanning probe microscope for high throughput metrology and inspection," Proc. SPIE **XXIX**, 94240O (2015).

[8]D. H. F. Y. Hui Xie, "Development of three-dimensional atomic Force microscope for sidewall structures imaging with controllable scanning density," IEEE/ASME Trans. Mechatron. **21**(1), 316 - 328 (2016).

[9]A. Majumdar, "Scanning thermal microscopy," Annu. Rev. Mater. Sci. **29**(1), 505-585 (1999).

[10]M. Despont, J. Brugger, U. Drechsler, U. Dürig, W. Häberle, M. Lutwyche, H. Rothuizen, R. Stutz, R. Widmer, G. Binnig, H. Rohrer, and P. Vettiger, "VLSI-NEMS chip for parallel AFM data storage," Sensors Actuators A Phys. **80**, 100–107 (2000).

[11]H. Sadeghian, R. W. Herfst, T. C. Van den Dool, W. E. Crowcombe, J. Winters, and G. F. I. J. Kramer, "High-throughput parallel SPM for metrology, defect, and mask inspection," Proc. SPIE **9231**, 92310B (2014).

[12]S. C. Minne, G. Yaralioglu, S. R. Manalis, J. D. Adams, J. Zesch, A. Atalar, and C. F. Quate, "Automated parallel high-speed atomic force microscopy," Appl. Phys. Lett. **72**, 2340-2342 (1998).

[13]H. C. Jo, H. J. Lim, S. J. Shin, J. H. Kim, Y. S. Kim, and S.-I. Park, "Scanning probe microscope with automatic probe replacement function." US Patent US7709791 B2, 2006.

[14]W.-S. Park, S.-H. Kim, and Y.-H. Kim, "Scanning probe microscope and method of operating the same." USA Patent US8925111 B1, June 2013.

[15]J. Kwon, J. Hong, Y. Kim, D. Lee, K. Lee, S. Lee, S. Park, "Atomic force microscope with improved scan accuracy, scan speed, and optical vision," Rev. Sci. Instrum. **74**, 4378-4383 (2003).

[16]R. Herfst, B. Dekker, G. Witvoet, W. Crowcombe, D. de Lange, and H. Sadeghian, "A miniaturized, high frequency mechanical scanner for high speed atomic force microscope using suspension on dynamically determined points," Rev. Sci. Instrum. **86**, 113703 (2015).

[17] H Sadeghian, N. B. Koster, and T. C. Van de Dool, "Introduction of a high throughput SPM for defect inspection and process control," SPIE Adv. Lithogr. **XXVII**, 868127, (2013).

[18]H Sadeghian, T. C. Van de Dool, W. E. Crowcombe, R. W. Herfst, J. Winters, G. F. I. J. Kramer, and N. B. Koster, "Parallel, miniaturized scanning probe microscope for defect inspection and review," SPIE Adv. Lithogr. **XXVIII**, 90501B (2014).

[19]R. Herfst, B. Dekker, G. Witvoet, W. Crowcombe, D. de Lange, and H. Sadeghian, "A miniaturized, high frequency mechanical scanner for high speed atomic

force microscope using suspension on dynamically determined points," Rev. Sci. Instrum. **86**, 113703 (2015).